\documentclass[11pt]{article}
\usepackage[psamsfonts]{amssymb}
\usepackage{amsmath}
\usepackage{color}

\author{H. Mohseni Sadjadi$^1$\footnote{mohsenisad@ut.ac.ir}\ \ and
Mubasher Jamil$^2$\footnote{mjamil@camp.nust.edu.pk}
\\ $^1${\small Department of Physics, University of Tehran, P.O.B.
14395-547}
\\ {\small Tehran 14399-55961, Iran}
\\$^2$ {\small Center for Advanced Mathematics and Physics,}\\
{\small National University of Sciences and Technology (NUST)}
\\ {\small Islamabad, Pakistan.}}

\title{Generalized second law of thermodynamics for FRW cosmology with
logarithmic correction}
\begin{document}
\maketitle

\begin{abstract}
\textbf{Abstract:} In the previous analyses in the literature about
the generalized second law (GSL) in an accelerated expanding
universe the usual area relation for the entropy, i.e. $S={A\over
4G}$, was used for the cosmological horizon entropy. But this
entropy relation may be modified due to thermal and quantum
fluctuations or corrections motivated by loop quantum gravity giving
rise to
\begin{equation}\nonumber
S={A\over 4}+\pi\alpha \ln({A\over 4})+\gamma,
\end{equation}
where $\alpha$ and $\gamma$ are some constants whose the values are
still in debate in the literature. Our aim is to study the
constraints that GSL puts on these parameters. Besides, we
investigate the conditions that the presence of such modified terms
in the entropy puts on other physical parameters the system such as
the temperature of dark energy via requiring the validity of GSL. In
our study we consider a spatially flat Friedman-Robertson-Walker and
assume that the universe is composed of several interacting
components (including dark energy). The model is investigated in the
context of thermal equilibrium and non-equilibrium situations. We
show that in a (super) accelerated universe GSL is valid whenever
$\alpha (<)>0$ leading to a (negative) positive contribution from
logarithmic correction to the entropy. In the case of super
acceleration the temperature of the dark energy is obtained to be
less or equal to the Hawking temperature.
\end{abstract}
\maketitle
\section{Introduction}

After the discovery of deeper relationships between gravity and
thermodynamics \cite{hawking}, it was realized that the notion of
temperature is not restricted to black hole horizons only. The
study of quantum field theory in any spacetime with a horizon
showed that all horizons have temperatures i.e. all horizons
behave like black body \cite{davi}. In this connection, Gibbons
and Hawking conjectured that entropy can be associated with the
cosmological horizons and other similar properties like
temperature and surface gravity as well \cite{gibbons}. In the
simplest cosmological model of de Sitter Universe, the temperature
goes like $T\sim\sqrt{\Lambda}$, where $\Lambda$ is a positive
cosmological constant. It gave the motivation that GSL can be
studied for the de Sitter spacetime and probably also for other
cosmological spacetimes \cite{wall}. In literature, GSL has been
widely discussed in the framework of different gravity theories
including Braneworld, Gauss-Bonnet modified gravity and $f(R)$
gravity \cite{gravity}. Also GSL is investigated in the presence
of dark energy \cite{energy} and black hole \cite{hole}. Jacobson
proved that Einstein field equation can be derived from the usual
Clausius relation, $dQ=TdS$, where $dQ$ is the energy exchange,
$T$ is temperature and $dS$ the change in entropy \cite{jacob}.
Later on Padmanabhan showed that Einstein's equations for a
spherically symmetric spacetime can be written in the form,
$TdS=dE+PdV$, near `any' horizon \cite{paddy2}.

The FRW Universe may contain several cosmic ingredients including
dark energy, dark matter and radiation. Astrophysical observations
suggest that the energy density of dark energy is the dominant
component of the total cosmic energy density. In a recent study
\cite{jamil1}, it has been proven that the GSL will be valid in
the cosmological scenario where dark energy interacts with both
dark matter and radiation. Also that the GSL is always and
generally valid, independently of the specific interaction form,
of the fluids equation of state parameters and of the background
(FRW) geometry. In that particular study, the authors assumed that
the FRW Universe is enclosed by the apparent horizon and all the
interacting components were in thermal equilibrium with the
apparent horizon (a trapped surface with vanishing expansion). We
think that the later two assumptions should be replaced with more
physically motivated ones: The temperature of cosmic ingredients
could be different from each other, for instance, radiation could
have higher temperature than `cold' dark matter and so on; also
the boundary of FRW Universe could be apparent horizon or future
event horizon (the distance that light can travel from now till
the end of time). The later horizon is best suitable as an
infrared cut-off for the Universe containing the holographic dark
energy \cite{jamil2}.

In the following analysis, we shall assume the FRW Universe to
contain several cosmic fluids interacting with each other and
exchanging energy densities. We also take into account the
possibility of thermal non equilibrium. We investigate the
generalized second law of thermodynamics in the context of FRW
cosmology by considering the logarithmic correction to the horizon
entropy. The outline of this paper is as follows: In the second
section, we discuss the cosmological model and calculate the
entropy rate of change of cosmic ingredients; next, we construct
the generalized second law with quantum-corrected entropy of the
horizon. The validity of this law forces the parameters the model
to satisfy some special conditions. This restricts the range of
free parameters of the model introduced in the literature. In the
next two subsections, we discuss GSL in thermal equilibrium and
non-equilibrium settings and finally we conclude the paper.

We use units $c=G=k_B=1$.

\section{The cosmological model}

We consider the spatially flat FRW spacetime
\begin{equation}\label{1}
ds^2=-dt^2+a^2(t)(dx^2+dy^2+dz^2)
\end{equation}
in comoving coordinates, where $a(t)$ is the scale factor. This
Universe is assumed to be composed of a n-components perfect fluid
(such as dark energy, dark matter, radiation and so on):
$\rho=\sum_{i=1}^n \rho_i$, $P=\sum_{i=1}^n P_i$, where $\rho$ and
$P$ are the total density of the energy and the pressure of the
Universe respectively. In this spacetime the Friedman equations
read
\begin{eqnarray}\label{2}
H^2&=&{8\pi\over 3}\rho,\nonumber \\
\dot{H}&=&-4\pi(P+\rho).
\end{eqnarray}
Here overdot represents differentiation with respect to the time
$t$. The first law of thermodynamics implies that each of the
components in the volume $V$ satisfies \cite{dav,arxiv}:
\begin{equation}\label{3}
dE_i=T_idS_i-P_idV.
\end{equation}
Here $E_i=\rho_iV$ is the energy, $T_i$ is the temperature and
$S_i$ is the entropy of $i$-th component of the perfect fluid.
From (3), we can write
\begin{equation}\label{4}
\dot{S_i}={1\over T_i}\left(
(P_i+\rho_i)\dot{V}+V\dot{\rho_i}\right).
\end{equation}
The continuity equation for each element is
\begin{equation}\label{5}
\dot{\rho_i}+3H(P_i+\rho_i)=Q_i,
\end{equation}
where $Q_i$ is an interaction term which can be an arbitrary
function of cosmological parameters like the Hubble parameter and
energy densities \cite{jamil}. This term allow the energy exchange
between the components of the perfect fluid and may alleviate the
coincidence problem. In our analysis, we proceed with a general
$Q_i$.

Divergenceless  of the energy momentum tensor leads to
$\dot{\rho}+3H(P+\rho)=0$, hence $\sum_{i=1}^n Q_i=0$. From (4) we
derive
\begin{equation}\label{6}
\dot{S_i}={VQ_i\over T_i}+{1\over T_i}(P_i+\rho_i)(\dot{V}-3HV).
\end{equation}
To proceed further we must specify $V$. In the literature there
are different choices for $V$ corresponding to different horizons
of the universe:

In cosmological models of accelerated universe, there are horizons
to which we can assign an entropy as a measure of information
behind them. So to study the entropy of the universe, besides the
entropy of the matter inclosed by the horizon, the horizon entropy
must also be taken into account. The most natural horizon of the
universe is the apparent horizon whose radius is $R_A=H^{-1}$. The
choice $V={4\pi\over 3H^3}$ has been adopted by many authors to
study of thermodynamics of the (accelerated) universe \cite{app}.

Another cosmological horizon which conceptually more resembles to
the black hole horizon is the future event horizon, whose radius,
$R_f$, is defined by (in the presence of the big rip at $t_s$,
$\infty$ must by replaced by $t_s$)
\begin{equation}\label{7}
R_f(t)=a(t)\int_t^{\infty}{dt'\over a(t')}.
\end{equation}
Due to the holographic description of dark energy proposed by
\cite{Li}, recently this horizon has attracted more attentions in
the study of thermodynamics of the universe. Only in a De Sitter
space time we have $R_A=R_f$. A detailed discussion about the
relation between $R_A$ and $R_f$ in an accelerated universe can be
found in \cite{bousso}.

For generality, in our study we consider both the choices: $R_A$
and $R_f$.

Using
\begin{equation}\label{9}
\dot{R_f}=HR_f-1,
\end{equation}
(4) becomes
\begin{equation}\label{10}
\dot{S}_i={VQ_i\over T_i}-4\pi R_f^2{P_i+\rho_i\over T_i}.
\end{equation}

Similarly, if we take the horizon  as the apparent horizon:
$R_A=H^{-1}$, we obtain
\begin{equation}\label{11}
\dot{S}_i={4\pi\over 3 H^3}{Q_i\over T_i}-4\pi\left({\dot{H}\over
H^4}+{1\over H^2}\right){P_i+\rho_i\over T_i}.
\end{equation}
Hence even in the thermal equilibrium: $T_i=T$ (where $T$ is the
temperature of the horizon) or in the absence of interactions:
$Q_i=0$, the total entropy of the components within the volume $V$
is not a constant. In the thermal equilibrium we have
$\dot{S}_{in}=R_f^2\dot{H}/T$ and $\dot{S}_{in}={1\over
T}({\dot{H}\over {H^2}}+1){\dot{H}\over {H^2}}$, corresponding to
(\ref{10}) and (\ref{11}) respectively. Note that we have
$\dot{S}_{in}<0$ for an accelerated expanding universe in
quintessence phase (i.e.  for $\dot{H}+H^2>0$ and $\dot{H}<0$).

\section{Generalized second law of thermodynamics with logarithmic correction}

To study the evolution of the entropy of the Universe, we take
into account the contribution of the entropy associated to the
horizon. Note that the expression of entropy is a quantity that is
derived from the theory of gravity under consideration. In
Einstein's gravity, the entropy of the horizon (both for black
holes and the FRW Universe) is proportional to the area of the
horizon, $S\propto A$. When the gravity theory is modified by
adding extra curvature terms in the action functional, it
ultimately modifies the entropy-area relation, for instance, in
$f(R)$ gravity, the entropy-area relation is, $S\propto f'(R)A$
\cite{mohseni}. In the context of loop quantum gravity, the
entropy-area relation can be expanded in an infinite series
expression. The leading order term in this expression is the
logarithmic correction term to entropy-area relation.
Mathematically, we have \cite{zhu}
\begin{eqnarray*}
S_h &=& \frac{A}{4 \hbar}+ \tilde{\alpha} \ln \frac{A}{4\hbar}
-\tilde{\alpha}_1\frac{4 \hbar}{A}-\tilde{\alpha}_2\frac{16
\hbar^2}{A^2}-\cdots\\&=&S_0+\tilde{\alpha}\ln
S_0-\sum_{i=1}^{\infty}\frac{\tilde{\alpha}_i}{S_0^i}.
\end{eqnarray*}
Here $S_0$ is the classical entropy of the black hole while the
second and higher order terms are non-logarithmic terms and are
called quantum corrections. Here $\tilde{\alpha}_i$ are finite
constants and $A$ is the area of the horizon. It is obvious from
the last expression that higher order terms are ignorable due to
smallness of $\hbar$ and only first order correction term is
relevant for the analysis. (From here onwards we shall fix
$\hbar=1$.) The issue of the value of $\tilde{\alpha}_i$'s is
highly disputatious. There are different interpretations found in
the literature. The prefactor of the logarithmic term,
$\tilde{\alpha}$, for example, is given to be $-3/2$ in \cite{KM},
and $-1/2$ in \cite{GM}. Similarly, some authors \cite{Hod} take
it to be a positive integer, while others find it even to be zero
\cite{Med}.

{\it{Note that besides the framework of loop quantum gravity, the
same result for the corrected entropy can be derived by
considering the effects of quantum \cite{quant}, and thermal
fluctuations around equilibrium \cite{therm}, charge and mass
fluctuations \cite{wei}}}.

In the following we only take into account the contribution of the
logarithmic correction such that \cite{wei}
\begin{equation}\label{12}
S_h={A\over 4}+\pi\alpha \ln({A\over 4})+\gamma,
\end{equation}
where $\pi\alpha\sim \mathcal{O}(1)$ and $\gamma$ are constants.
This logarithmic term also appears in a model of entropic cosmology
which unifies the inflation and late time acceleration \cite{cai}.
It is very interesting if one can determine the coefficient $\alpha$
in front of log correction term by observational constraints. In
this connection, the same study showed that the coefficient might be
extremely large, of order $10^{16}$, due to current cosmological
constraints which inevitably brought a fine tuning problem to
entropy corrected models \cite{cai}. Hence the logarithmic
correction might play a huge role in cosmology. From (\ref{11}) and
(\ref{12}), the time evolution of the total entropy (or the
generalized entropy) defined by $\dot S=\dot{S}_h+\dot{S}_{in}$, for
$R=R_A$ is
\begin{equation}\label{13}
\dot{S}={4\pi\over 3H^3}\sum_{i=1}^{n}{Q_i\over
T_i}-4\pi\left({\dot{H}\over H^4}+{1\over
H^2}\right)\sum_{i=1}^{n}{P_i+\rho_i\over T_i}-2\pi{\dot{H}\over
{H^3}}(1+\alpha H^2),
\end{equation}
while for $R=R_f$, the total entropy is obtained as
\begin{equation}\label{14}
\dot{S}={4\pi R_f^3\over 3}\sum_{i=1}^{n}{Q_i\over T_i}-4\pi
R_f^2\sum_{i=1}^{n}{P_i+\rho_i\over
T_i}+2\pi(HR_f-1)\Big(R_f+{\alpha\over R_f}\Big).
\end{equation}
The GSL requires $\dot{S}\geq0$, i.e. the sum of the entropies of
the perfect fluids inside the horizon and the entropy attributed
to the horizon is a non decreasing function of the comoving time.
In the following we discuss the validity of this law specially in
the presence of dark energy.

\subsection{GSL in thermal equilibrium}

In the thermal equilibrium, i.e. $\forall i: T_i=T$, (\ref{13})
becomes
\begin{equation}\label{15}
\dot{S}={\dot{H}\over T}\left({\dot{H}\over H^4}+{1\over
H^2}\right)-2\pi{\dot{H}\over H^3}(1+\alpha H^2).
\end{equation}
When the horizon is the apparent horizon we take the temperature
as the Hawking temperature $T={H\over 2\pi}$ and GSL for
(\ref{13}) reduces to
\begin{equation}\label{17}
\dot{S}={2\pi\dot{H}\over H^3}\left({\dot{H}\over H^2}-\alpha
H^2\right)\geq 0.
\end{equation}
Hence GSL is always satisfied in the absence of the correction
term. Note that in a de Sitter spacetime, i.e. $\dot{H}=0$, the
expansion is isentropic : $\dot{S}=0$.

(\ref{17}) puts some constraint on $\alpha$. As an illustration
consider the case of small perturbations around de Sitter space
(quasi de Sitter spacetime):
\begin{equation}\label{18}
H=H_0+H_0^2\epsilon t+\mathcal{O}(\epsilon^2),
\,\,\epsilon:={\dot{H}\over H^2}, \,\, |\epsilon|\ll 1,
\dot{\epsilon}=\mathcal{O}(\epsilon^2).
\end{equation}
When $\dot{H}>0$, i.e. in a super accelerated universe, the GSL is
satisfied only when $\alpha <{\epsilon \over H^2}$. In the units
where $G\neq 1$ this can be rewritten as $8\pi\alpha
<{\epsilon}{m_P^2 \over H^2}$, where $m_P^2={8\pi\over G}$, and
$m_P$ is the Planck mass. Note that, as $\epsilon \ll 1$, the left
hand side of this inequality may be a number of order unity.
$\alpha<0$ is enough (although not necessary) condition to satisfy
GSL in this case for all small positive values of $\epsilon$. In
the same way when $\dot{H}< 0$, e.g. in dark energy dominated era
and when the universe is in the quintessence phase, the GSL is
satisfied for $\alpha >{\epsilon \over H^2}$ or $8\pi\alpha
>{\epsilon}{m_P^2 \over H^2}$. Here, $\alpha>0$ is enough
condition for validity of GSL. Note that theses inequalities do
not determine the precise value of $\alpha$, and corresponding to
the model proposed (e.g. for inflation or late time accelerated
expansion of the universe where we can consider a quasi de Sitter
space time), only give us the upper or lower bound of $\alpha$.
Note that as our results are general and independent of the
explicit form of dark energy or inflaton,  we have not determined
numerically theses bounds which are model dependent.

For the future event horizon, and in the absence of a well defined
temperature, we assume that $T$ is proportional to the Hawking
temperature \cite{dav}
\begin{equation}\label{16}
T={bH\over 2\pi}.
\end{equation}
GSL requires
\begin{equation}\label{19}
\dot{S}=2\pi{\dot{H}\over
bH}R_f^2+2\pi(HR_f-1)\Big(R_f+{\alpha\over R_f}\Big)\geq 0.
\end{equation}
This indicates that
\begin{equation}\label{20}
{d \over dt}\ln (R_fH^{1\over b}e^{{-\alpha\over 2R_f^2}})\geq 0.
\end{equation}
If the expansion is isentropic then $H=\lambda R_f^{-b}e^{\alpha
b\over 2R_f^2}$, where $\lambda$ is a positive constant. By using
(\ref{9}) we conclude that the future event horizon satisfies
\begin{equation}\label{21}
\dot{R_f}=\lambda R_f^{1-b}e^{\alpha b\over 2R_f^2}-1.
\end{equation}
If the Universe remains in the phantom phase then $\dot{R_f}<0$
\cite{dav}. This implies $\lambda<R_f^{b-1}e^{-\alpha b\over
2R_f}$. But near the big rip we have  $\lim_{t\to t_s}R_f=0$,
hence positivity of $\lambda$ requires $\alpha<0$. In this case we
do not need to have $b>1$, which is true only when $\alpha=0$.

To see more explicitly how the validity of GSL depends on the
value of $\alpha$, let us consider a model, discussed in the
literature (see \cite{energy} and references therein),
corresponding to a super accelerated universe with a big rip at
the finite time $t_s$, near which $H$ becomes very large. This
phantom dominated universe of pole-like type is described by the
scale factor
\begin{equation}\label{22}
a(t)=a_0(t_s-t)^{-n},
\end{equation}
where $a_0$ and $n$ are positive constants.  The Hubble parameter
and the future event horizon are given by $H={n\over {t_s-t}}$ and
$R_f={{t_s-t}\over {n+1}}$ respectively. Hence (\ref{19}) reduces
to
\begin{equation}\label{23}
(b-1)(t_s-t)^2+b\alpha(n+1)^2\leq 0.
\end{equation}
For $\alpha=0$, GSL is satisfied provided that $b\leq 1$ (the
temperature is less than the Hawking temperature). In the presence
of the correction term, for $b>1$ (i.e. the temperature is greater
than the Hawking temperature), GSL is satisfied when
\begin{equation}\label{24}
(t_s-t)^2\leq {b\alpha(n+1)^2\over 1-b}.
\end{equation}
In this case we must have $\alpha<0$. For temperatures less than
the Hawking temperature, the GSL is valid when
\begin{equation}\label{25}
(t_s-t)^2\geq {b\alpha(n+1)^2\over 1-b},
\end{equation}
which is  true for all $t$, provided that $\alpha<0$.

\subsection{GSL with thermal non-equilibrium}

If we leave the thermal equilibrium condition, the problem of
investigating the validity of the GSL becomes more complicated.
However to get an insight about what happens in this case, let us
consider a simple model. Assume that the Universe is dominated by
two subsets with different temperatures: $\rho=\rho_1+\rho_2$,
such that each of them satisfies the continuity equation:
$\dot{\rho_j}+3H(P_j+\rho_j)=0$,\,\,\, $j=1,2$ and evolves with
its own temperature. Note that in this model $\rho_1$ and $\rho_2$
are consisted of ingredients {\it in thermal equilibrium (in each
subset)}, which although don't interact with the elements of the
other subset, can interact with each other. Hence in this
situation using (\ref{13}), (\ref{14}) one can show that GSL
requires
\begin{eqnarray}\label{26}
\dot{S}&=&\left({\dot{H}\over H^4}+{1\over
H^2}\right)\left(-4\pi(\rho_2+P_2)\left({1\over T_2}-{1\over
T_1}\right)+{\dot{H}\over T_1}\right)\nonumber
\\
&&-2\pi{\dot{H}\over H^3}\left(1+\alpha H^2\right)\geq 0,
\end{eqnarray}
and
\begin{equation}\label{27}
\dot{S}=-4\pi R_f^2\left((P_2+\rho_2)({1\over T_2}-{1\over
T_1})\right)+{R_f^2\dot{H}\over T_1}+2\pi(HR_f-1)(R_f+{\alpha\over
R_f})\geq 0,
\end{equation}
respectively. These inequalities written in this special form
allow us to study the validity of the GSL by knowing the EoS
parameter of only one of the sectors. If we take the first sector
as the dark energy perfect fluid whose temperature is assumed to
be $T_1={bH\over 2\pi}$ and the other sector as a barotropic
perfect fluid, $P_2=w_2\rho_2$, (\ref{26}) leads to
\begin{eqnarray}\label{28}
\dot{S}&=&\Big(-4\pi(w_2+1)\rho_{02}a^{-3(w_2+1)}\Big({1\over
T_2}-{2\pi\over bH}\Big)+{2\pi\dot{H}\over
bH}\Big)\nonumber\\&&\times\Big({\dot{H}\over H^4}+{1\over
H^2}\Big)-2\pi{\dot{H}\over H^3}\Big(1+\alpha H^2\Big)\geq 0,
\end{eqnarray}
where $\rho_{02}=\rho_2(a=1)$ is a constant and we have used the
continuity equation to write $\rho_2=\rho_{02}a^{-3(w_2+1)}$. In
the same way (\ref{27}) casts to
\begin{eqnarray}\label{29}
\dot{S}&=&-4\pi(1+w_2)\rho_{02}R_f^2a^{-3(w_2+1)}\left( {1\over
T_2}-{2\pi\over bH}\right)+{2\pi\dot{H} R_f^2
\over bH}\nonumber \\
&+&2\pi(HR_f-1)(R_f+{\alpha\over R_f})\geq 0.
\end{eqnarray}

The non equilibrium situation changes drastically the previous
necessary conditions for validity of GSL. For example, as it is
obvious from (\ref{26}) or (\ref{28}), even in a de Sitter
spacetime the expansion is no more isentropic and the GSL is valid
provided that
\begin{equation}\label{30}
(\rho_2+P_2)\left({1\over T_2}-{1\over
T_1}\right)=(\rho_1+P_1)\left({1\over T_1}-{1\over T_2}\right)\leq
0.\end{equation} Hence GSL implies that the temperature of a
sector which is in quintessence (phantom) phase is greater (less)
than the other component.

For a known $T_2$ (in terms of the scale factor), the inequalities
(\ref{28}) and (\ref{29}) determine the time derivative of the
total entropy in terms of the scale factor, its time derivatives,
and its time integral. So in principle they specify the needed
conditions required for the GSL to be true. As a choice one can
take the second sector as an ordinary matter like radiation, whose
temperature may obtained as $T_2=T_{02}a^{-3w_2}$, where
$T_{02}=T(a=1)$. This follows from the equation
\begin{equation}
{\dot{T_i}\over T_i}=w_i{Q_i-3H(P_i+\rho_i)\over P_i+\rho_i}.
\end{equation}
For non-interacting radiation this gives the well known result:
$T_2=T_{02}a^{-1}$. Another choice is to take the temperature of
the second sector like the first one as $T_2={b_2H\over 2\pi}$
where $b_2\neq b$ is a positive constant.

Following (\ref{29}), for $T_2=T_{02}a^{-3w_2}$, GSL is true
whenever
\begin{equation}\label{31}
A'(t_s-t)^{3(n+1)}+B(t_s-t)^{3n(w_2+1)+4}+C(t_s-t)^2\geq \alpha,
\end{equation}
where
\begin{equation*}
A'=-{2(1+w_2)\rho_{02}\over T_{02}(n+1)^2a_0^3},\ \ \ B={4\pi
(1+w_2)\rho_{02}\over bn(n+1)^2a_0^{3(1+w_2)}},\ \  C={1-b\over
b(n+1)^2}.
\end{equation*}
Note that as $A'$ and $B$ have different signs, even in the case
$\alpha<0$ and $b<1$, GSL may not be satisfied for some times. Now
let us examine the other choice, $T_2={b_2H\over 2\pi}$. In this
case GSL is valid when
\begin{equation}\label{32}
F (t_s-t)^{3n(w_2+1)+2}+C(t_s-t)^2\geq \alpha,
\end{equation}
where
\[
F= -{4\pi n (1+w_2)\rho_{02}\over (n+1)^2 a_0^{3(w_2+1)}}
\left({1\over b_2}-{1\over b}\right).
\]
If the dark energy temperature is greater than the Hawking
temperature, $b>1$, then $C<0$ and for $\alpha>0$, GSL is
satisfied for $F>0$. If the dark energy temperature is less than
the Hawking temperature $b<1$, the GSL is satisfied {\it always}
provided that $\alpha<0$ and $F>0$. Note that $F>0$ leads to
$b_2>(<)b$ for quintessence (phantom) like $w_2$.

At the end we would like to note that if one considers an
interaction between the components which have different
temperatures, then the first terms in (\ref{13}) and (\ref{14})
must be taken into account too, and the problem becomes very
complicated. This is due to the fact that for a general
interaction obtaining an expression for the energy densities
$\rho_i$ or temperature in terms of the scale factor is not
straightforward.
\\
\section{Conclusion}
We considered the FRW cosmological spacetime
\textcolor{red}{(see(\ref{1}))} composed of interacting components
and discussed the generalized second law (GSL) of thermodynamics by
considering logarithmic correction to the horizon entropy:
\begin{equation}\nonumber
S={A\over 4}+\pi\alpha \ln({A\over 4})+\gamma.
\end{equation}
$\alpha$ and $\gamma$ are constant, which their values are in debate
in the literature (see the discussion in the beginning of the third
section and after (\ref{12})). We tried to examine this modified
entropy for cosmological horizons to see whether the GSL can
restrict the value of the coefficient of the logarithmic
leading-order correction, $\alpha$, and also the consequence of such
correction in the thermodynamical parameters of an accelerated
universe such as the temperature of dark energy.

For the case of generality we performed our analysis by focusing on
both apparent and future event horizons and reexpressed GSL in terms
of the parameters of the model (see (\ref{13}) and (\ref{14})). We
showed that, although by considering the usual area-entropy relation
the GSL holds for the apparent horizon, the logarithmic correction
term may destroy the validity of $GSL$ (see(\ref{17})). In a quasi
de Sitter space time which is a model corresponding to the expansion
of the universe in the early universe or at late time
(see(\ref{18})), it was shown that that (negativity) positivity of
$\alpha$ in a (super) accelerated universe is enough condition for
validity of GSL leading to a negative (positive) contribution from
correction to the entropy. The constraint coming from GSL which are
stated as inequalities (depending on the model proposed for the
evolution of the universe), only enabled us to determine the upper
or lower bounds of $\alpha$.

We performed similar analysis for the future event horizon via some
examples concerning the pole like type expansion and big rip
(see(\ref{22})) where the Hubble parameter may become very large
with respect to the present time. We showed that in order that GSL
be satisfied the temperature of dark energy must be less than the
Hawking temperature, and $\alpha$ must be negative: therefore the
correction decreases the horizon entropy.

A brief analysis was also done for a universe dominated by two non
interacting components with different temperatures (although each
components may contain interacting ingredients in thermal
equilibrium)(see(\ref{26}) and (\ref{27})). In the case of future
event horizon we verified that GSL hold always provided that
$\alpha<0$, so as before the horizon entropy is decreased by
logarithmic correction.

Note that as the GSL deals with the time derivative of the total
entropy, our analysis gives not any information about $\gamma$.

\subsubsection*{Acknowledgment}
We would like to thank J. Bekenstein for useful comments on this
paper.


\begin{thebibliography}{99}

\bibitem{hawking} J.D. Bekenstein, Nuovo Cim. Lett. 4 (1972) 737;\\ S.W. Hawking,
Commun. Math. Phys. 43 (1975) 199;\\ J.M. Bardeen, B. Carter and S.W. Hawking, Comm. Math. Phys. 31 (1973) 161;\\
 J.D. Bekenstein, Phys. Rev. D 7 (1973) 2333.

\bibitem{davi} S.A. Fulling, Phys. Rev. D 7 (1973) 2850;\\ P.C.W.
Davis, J. Phys. A 8 (1975) 609.
\bibitem{gibbons} G.W. Gibbons and S.W. Hawking, Phys. Rev. D 15 (1977)
2738.
\bibitem{wall} T.M. Davies et al, Class. Quant. Grav. 20 (2003)
2753;\\ K Karami, S Ghaffari and M M Soltanzadeh, Class. Quant.
Grav. 27 (2010) 205021.
\bibitem{gravity} M. Akbar, Int. J. Theor. Phys. 48 (2009) 2665;\\
M. Akbar, Chin. Phys. Lett. 25 (2008) 4199.

\bibitem{energy} H.M. Sadjadi, Phys. Rev. D 73 (2006) 063525.

\bibitem{hole} G. Izquierdo and D. Pavon, Phys. Lett. B
639 (2006) 1;\\ A. Hosoya et al, Phys. Rev. D 63 (2001) 104008;\\
H.M. Sadjadi, Phys. Lett. B 645 (2007) 108.
\bibitem{jacob} T. Jacobson, Phys. Rev. Lett. 75 (1995) 1260.
\bibitem{paddy2} T. Padmanabhan, Class. Quant. Grav. 19 (2002) 5387.


\bibitem{jamil1} M. Jamil, E.N. Saridakis and M.R. Setare, Phys. Rev. D 81 (2010) 023007.

\bibitem{jamil} H.M. Sadjadi and M. Honardoost,
Phys. Lett. B 647 (2007) 231;\\ H.M. Sadjadi and M. Alimohammadi,
Phys. Rev. D 74 (2006) 103007;\\ H.M. Sadjadi, arXiv:0904.1349v3
[gr-qc];\\ H.M. Sadjadi, arXiv:0909.1002v1 [gr-qc].

\bibitem{jamil2}
M. Jamil and M.U. Farooq, Int. J. Theor. Phys. 49 (2010) 42;\\
H.M. Sadjadi, JCAP 02(2007) 026;\\ H.M. Sadjadi, Eur. Phys. J. C
62 (2009) 419.

\bibitem{dav} P.C.W. Davies, Class. Quant. Grav. 4 (1987 ) L225;\\
H. M. Sadjadi, Phys. Rev. D 73 (2006) 063525.

\bibitem{arxiv} T. Clifton and J.D. Barrow, Phys. Rev. D 75 (2007)
043515.
\bibitem{app} Y. Gong,  and A.  Wang ,  Phys. Rev. Lett. {\bf 99}, 211301
(2007).

\bibitem{Li}M. Li, Phys. Lett. B 603, 1 (2004).
\bibitem{bousso}R. Bousso, Phys. Rev. D 71, 064024 (2005).

\bibitem{mohseni} R.M. Wald, Phys. Rev. D 48 (1993) 3427.

\bibitem{zhu}  M. Jamil and M. U. Farooq, JCAP 03 (2010) 001;\\ R. Banerjee and S. K.
Modak, JHEP 0905 (2009) 063;\\ S. K. Modak, Phys. Lett. B 671
(2009) 167.

\bibitem{KM} R.K. Kaul and P. Majumdar, Phys. Rev. Lett. 84 (2000) 5255.
\bibitem{GM} A. Ghosh and P. Mitra, Phys. Rev. D 71 (2005) 027502.
\bibitem{Hod} S. Hod, Class. Quant. Grav. 21 (2004) L97.
\bibitem{Med} A.J.M. Medved, Class. Quant. Grav. 22 (2005) 133.

\bibitem{quant}R. B. Mannand, S. N. Solodukhin, Nucl.
Phys. B 523 (1998) 293.

\bibitem{therm}S. Das, P. Majumdar and R. K. Bhaduri, Class.
Quant. Grav. 19 (2002) 2355;\\ A. Chatterjee and P. Majumdar,
arXiv:gr-qc/030330.

\bibitem{wei} H. Wei, Commun. Theor. Phys. 52 (2009) 743.
\bibitem{cai} Y.F. Cai, J. Liu, H. Li, Phys. Lett. B 690 (2010)
213;\\ M. R. Setare and M. Jamil, arXiv:1011.0875 [gr-qc].
\end{thebibliography}
\end{document}